\documentclass[letterpaper, 10pt, conference]{ieeeconf}
\IEEEoverridecommandlockouts                              % This command is only
\usepackage[dvipdfm]{graphicx}
\usepackage[usenames,dvipsnames]{color}
\usepackage{amsmath}%
\usepackage{amsfonts}%
\usepackage{amssymb}%
\usepackage{graphicx}%
\usepackage{multicol}%
\usepackage{verbatim}%
\usepackage{enumerate}%

\usepackage[left=0.75in,top=1in,right=0.75in,bottom=0.75in]{geometry}

\title{
Cognitive Radio Transmission Strategies for Primary Erasure Channels
}

\author{
\authorblockN{Ahmed El-Samadony, Mohammed Nafie and Ahmed Sultan}
\authorblockA{Wireless Intelligent Networks
Center (WINC)\\
Nile University, Cairo, Egypt\\
Email: ahmed.elsamadony@nileu.edu.eg, mnafie@nileuniversity.edu.eg, salatino@stanfordalumni.org}}

\begin{document}

\setlength{\pdfpageheight}{\paperheight}
\setlength{\pdfpagewidth}{\paperwidth}

% You may need to change the horizontal offset to do what you
% want.  Setting \hoffset to a negative value moves all printed
% material to the left on all pages; setting it to a positive value
% moves all printed material to the right on all pages; not setting
% it keeps all printed material in it's default position.  \voffset
% is the vertical offset: use negative value for up; don't set if
% you want to use default position; use positive for down.
% \hoffset = -0.2truein
% \voffset = -0.2truein

\maketitle
\thispagestyle{empty}
\pagestyle{empty}

%%%%%%%%%%%%%%%%%%%%%%%%%%%%%%%%%%%%%%%%%%%%%%%%%%%%%%%%%%%%%%%%%%%%%%%%%%%%%%%%
\begin{abstract}

A fundamental problem in cognitive radio systems is that the cognitive radio is ignorant of the primary channel state and the interference it inflicts on the primary license holder. In this paper we assume that the primary transmitter
sends packets across an erasure channel and the primary receiver employs ACK/NAK feedback (ARQ) to retransmit erased
packets. The cognitive radio can eavesdrop on the primary's ARQs. Assuming the primary channel states follow a Markov
chain, this feedback gives the cognitive radio an indication of the primary link quality. Based on the ACK/NACK received, we devise optimal transmission strategies for the cognitive radio so as to maximize a weighted sum of primary and secondary throughput. The actual weight used during network operation is determined by the degree of protection afforded to the primary link. We study a two-state model where we characterize a scheme that spans the boundary of the primary-secondary rate region. Moreover, we study a three-state model where we derive the optimal strategy using dynamic programming. We also show via simulations that our optimal strategies achieve gains over the simple greedy algorithm for a range of primary channel parameters.

\end{abstract}

%%%%%%%%%%%%%%%%%%%%%%%%%%%%%%%%%%%%%%%%%%%%%%%%%%%%%%%%%%%%%%%%%%%%%%%%%%%%%%%%
\section{Introduction}
Cognitive radio technology is a solution to the problem of spectrum under-utilization caused mainly by static spectrum allocation. In cognitive radio networks, the licensed users coexist with cognitive users, also known as the secondary or unlicensed users. The secondary users attempt to utilize the resources unused by the primary users adopting procedures that aim at protecting the primary network from service interruption and interference.

There has been interest in schemes that make use of the feedback of the primary link to predict the behavior of the primary user in the future and, in the case of primary channel temporal correlation, to gain knowledge about the channel between primary transmitter and receiver (e.g., \cite{Bits through ARQs}, \cite{Cognitive Power Control} and \cite{Exploiting Hidden Power-Feedback}). In \cite{Bits through ARQs}, the secondary user observes the automatic repeat request (ARQ) feedback from the primary receiver. The ARQs reflect the primary user achieved packet rate. The cognitive radio's objective is to maximize secondary throughput under the constraint of guaranteeing a certain packet rate for primary user. The main difference between our work presented in this paper and \cite{Bits through ARQs} is that in \cite{Bits through ARQs} there is no use of the possible channel correlation across time, whereas we assume that the primary channel state follows a Markov chain. The cognitive transmitter can hence exploit the ARQs to predict the primary channel state during the next transmission phase. In \cite{Cognitive Power Control}, assuming a temporally correlated channel between the primary transmitter and receiver, the cognitive transmit power is adjusted based on primary channel state information (CSI) feedback. A real-time fading channel model is assumed rather than a binary erasure channel as we consider and discuss below. However, the computation of the optimal procedure in \cite{Cognitive Power Control} is computationally prohibitive.

There has been a series of recent work on cognitive MAC for opportunistic spectrum, e.g., \cite{Zhao}, \cite{Opportunistic Spectrum Access}, and \cite{Ding}. In \cite{Zhao}, an analytical framework for opportunistic spectrum access is developed on the basis of Partially Observable Markov Decision Processes (POMDP). The framework of POMDP is needed given the uncertainties about the quality of the primary link, and about primary activity as a result of sensing errors. In \cite{Zhao} a slotted primary network was considered, where primary activity remains fixed over the duration of a slot and switches between idle and active states according to a two-state Markovian process. The channel between the primary transmitter and receiver is not considered, and the feedback used to predict the channel availability is provided by the secondary receiver. In \cite{Opportunistic Spectrum Access}, the work in \cite{Zhao} is expanded to account for energy consumption and spectrum sensing duration optimization. In \cite{Ding}, the authors focus on the ARQ messages used in primary data-link-control and which are overheard by the secondary transmitter. Exploiting the primary feedback signals, the secondary terminal can optimize its access policy by assessing primary reception quality. The primary channel is assumed to be of fixed quality resulting in two fixed and known packet error rates corresponding to the presence and absence of secondary transmissions.

In this paper, we consider a primary transmitter that is always on. It sends a packet at each time slot, which has a fixed duration, and receives an ACK or NACK feedback from its receiver. The feedback is received correctly by both the primary and secondary transmitters. The channel between the primary transmitter and receiver is modeled as a Markov process with a finite number of states that determine the probability of correct reception. In this paper, we study primary link models with two and three states. The state of the channel does not change over a slot. The channel may switch states at the beginning of each slot according to the transition probabilities of the Markov process. The cognitive user exploits the ACK/NACK feedback from the primary receiver to predict the quality of the primary link. At the beginning of each time slot, the secondary user decides whether to remain silent and listen to primary feedback, or to carry out transmission. The objective is to maximize the weighted sum throughput of both the primary and secondary links.

Our contributions in this paper are as follows. For the two-state case, we find a closed form expression of the weighted sum throughput, and find the strategy that maximizes this throughput for any weight. Changing the weight spans the boundary of the  primary-secondary rate region. For the three-state case, we model the problem as a dynamic programming problem, and employ Bellman's equation \cite{Dynamic Programming} to arrive at the optimal strategy. In this paper we focus on the single channel case, but our scheme can be readily extended to the multiple channel case.

One of the advantages of our scheme is that the ARQ feedback can capture the temporal correlation in the channel. The cognitive user can access the primary channel in both cases, when the primary channel quality is relatively high (primary can transmit successfully regardless of cognitive user activity) and when its quality is very low (primary transmission fails whether secondary is active or not). This advantage cannot be captured in schemes employing spectrum sensing only.

The paper is organized as follows. The two-state system model and assumptions are described in Section \ref{sec:Two-state System model} where we find a closed form solution for the optimal throughput for primary and secondary networks. In Section \ref{sec:Three-state system model}, the three-state system model is examined. Numerical results are presented in Section \ref{sec:Simulation Results}. Our work is concluded in Section \ref{sec:Conclusion}.
\section{Two-state System model}
\label{sec:Two-state System model}
\begin{figure}[!ht]
\centering
  \includegraphics[width=0.9\columnwidth]{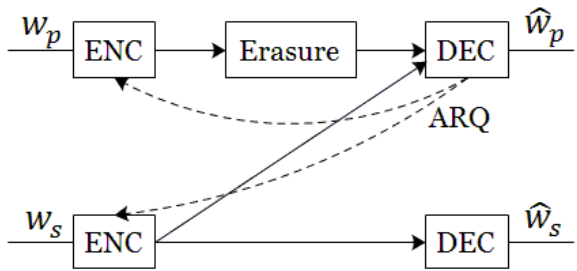}\\
  \caption{Z-interference erasure channel model. The secondary transmitter, when active, causes interference on the primary receiver. The secondary receiver, on the other hand, is shadowed from the primary transmitter, thereby suffering no interference from it. ENC is the channel encoder, whereas DEC is the receive decoder.}\label{fig:system_model2}
\end{figure}

Our proposed model assumes that we have one primary link and one secondary link. An illustration of the setup is provided in Figure \ref{fig:system_model2}. We are concerned with Z-interference channel model \cite{Z channel} where the interference from the primary transmitter on the secondary receiver is ignored. The Z-interference channel models important applications such as the interference caused by macro-cell users on femto-cell receivers, which is known in the literature as the ``loud neighbor'' problem. In our context, the primary terminals may be close to one another and use small transmission power, whereas the cognitive terminals may be far from one another and use high power for communication causing considerable interference on the primary link.

\begin{figure}[!ht]
	\centering
  \includegraphics[width=0.9\columnwidth]{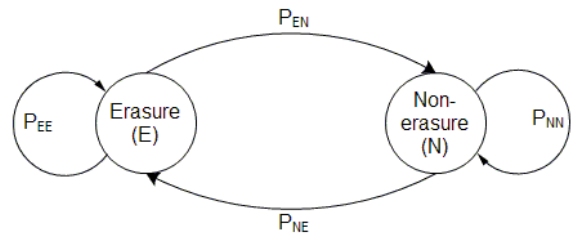}\\
  \caption{Two-state Markov model.}\label{fig:states2}
\end{figure}

We assume that the activity factor of the primary link is unity i.e., the primary transmitter sends a packet at each time slot. The primary link follows a two-state Markov chain. The primary link is either in an erasure (E) state or a non-erasure (N) state during each time slot. It switches states from one time slot to the next according to a Markovian process as shown in Figure \ref{fig:states2}. The process is specified by two parameters $P_{\rm EE}$ and $P_{\rm NE}$, where $P_{\rm EE}$ is the probability that the primary network is in erasure in the next time slot given that it is in erasure state in the current slot, and $P_{\rm NE}$ is the probability that the primary network is in erasure in the next time slot given that it is in non-erasure state in the current slot. The transition probabilities of the Markov chain are known a priori. The transition matrix $P$ which includes the transition probabilities is given by
\\
\[
P =
\left[ {\begin{array}{cc}
 P_{\rm EE} & P_{\rm EN}  \\
 P_{\rm NE} & P_{\rm NN}  \\
 \end{array} } \right]
\]
Hence, the stationary probabilities of being in erasure and non-erasure for primary network are $P(E)=\frac{P_{\rm NE}}{P_{\rm NE}+P_{\rm EN}}$ and $P(N)=\frac{P_{\rm EN}}{P_{\rm NE}+P_{\rm EN}}$, respectively.

The erasure state causes the primary transmission to fail, while the non-erasure state results in successful packet delivery to the primary receiver only when there is no interference from the secondary transmitter. That is, if the cognitive user decides to transmit in the non-erasure state, its transmission causes the erasure of the primary packet.

The cognitive radio can eavesdrop on the primary ARQ through which the secondary transmitter can detect the state of the primary link and, consequently, know the erasure probability of the next state using the transition probabilities. However, if the cognitive radio decides to transmit at a certain time slot, it causes the primary packet to be erased at this time slot. The secondary user then overhears a negative acknowledgment (NACK) from the primary receiver no matter what the state of the primary channel is. This means that when the cognitive user transmits, it becomes unaware of the primary link state.

Our objective is to choose the transmission strategy that maximizes the weighted sum throughput $Thr$ given by
\begin{equation}
Thr=wR_{\rm p}+\left(1-w\right)R_{\rm s}
\end{equation}
where $R_{\rm p}$ and $R_{\rm s}$ are the mean primary and secondary throughput, respectively, and $0\leq w \leq 1$ is a weighting factor that determines the relative importance of the two rates. In order to protect the primary user from interference and service interruption, parameter $w$ can be chosen close to one. The optimization problem has an exploration-exploitation tradeoff aspect. The tradeoff is between the cognitive user activity which maximizes the secondary user throughput, and cognitive user silence which gives the secondary user knowledge about the channel state information of the primary link through the ARQ feedback.

The primary reward, if the primary succeeds to transmit one packet through the binary erasure channel,
is $r_{\rm p}$. The secondary reward, if the secondary decide to transmit a packet, is $r_s$. Note that we can take account of any possible packet loss in the secondary channel when we calculate the value of $r_s$. The expected primary throughput at time slot $t$ estimated by the secondary transmitter is given by $r_{\rm p}\left(1-p_t\right)$, where $p_t$ is the secondary belief that the channel is in erasure at time $t$. The belief is updated from one time slot to another according to the following.
\\
\\
\small
$p_t = \left\{ \begin{array}{cc}
{P_{\rm EE}} & \mbox{if current state is erasure}
 \\{P_{\rm NE}} & \mbox{if current state is non-erasure}  \\
{p_{t-1}P_{\rm EE}+(1-p_{t-1})P_{\rm NE}} & \mbox{if current state is unknown}
\end{array}\right.$
\normalsize
\\
\\Note that the third possibility occurs when the secondary transmits. The expected secondary throughput at time slot $t$ is given by $r_{\rm s}I_t$ where $I_t$ is an indicator function given by\\

$I_t = \left\{ \begin{array}{rcl}
{0} & \mbox{if the secondary is silent at time slot $t$}
 \\ {1} & \mbox{if the secondary is active at time slot $t$}
\end{array}\right.$

%\\We analyze this problem using the framework of POMDP and prove its optimality by comparing results with the
%optimal control policy.

\subsection{Throughput Maximizing Scheme}

We assume that $P_{\rm EE}>P_{\rm NE}$ making the belief $p_t$ a monotonic function with time as long as
 the secondary user is transmitting \cite{Restless Bandit Problems}. This can be readily seen by solving the first order difference equation governing the evolution of $p_t$ to obtain
\begin{equation}
 p_t=(P_{\rm EE}-P_{\rm NE})^{k}p_{t-k}+\left[1-(P_{\rm EE}-P_{\rm NE})^{k}\right]P\left(E\right)
\end{equation}
where $p_{t-k}$ is the probability of being in erasure at time slot $t-k$ and the secondary is transmitting for $k$ consecutive slots, and $P(E)$ is the steady state probability as mentioned above. It is clear that if $P_{\rm EE}-P_{\rm NE}>0$, the belief $p_t$ is a monotonic function with time, otherwise the term $(P_{\rm EE}-P_{\rm NE})^{k}$ oscillates between positive and negative values.

If $wr_{\rm p}\left(1-P_{\rm EE}\right)>(1-w)r_{\rm s}$, the optimal secondary strategy is to listen always because the inequality also implies $wr_{\rm p}\left(1-P_{\rm NE}\right)>(1-w)r_{\rm s}$ which means that regardless of the actual system state, whether it is $E$ or $N$, the expected primary throughput is greater than the expected secondary throughput. Similarly, if $wr_{\rm p}\left(1-P_{\rm NE}\right)<(1-w)r_{\rm s}$, the optimal secondary strategy is to transmit always. For any other condition, the optimal secondary strategy is as follows. The secondary transmitter listens as long as an ACK is received because $wr_{\rm p}\left(1-P_{\rm NE}\right)>(1-w)r_{\rm s}$ and in that case maximizing the throughput in the next time slot is optimal since we do not affect future decisions as the secondary will make use of the knowledge of the next state while it is silent. Once a NACK is received, the secondary transmits $M$ consecutive packets. Thus, the maximization problem is equivalent to choosing optimal secondary consecutive transmitted packets $M$ that maximizes the weighted sum throughput.

\begin{figure}[!ht]
\centering
  \includegraphics[width=0.8\columnwidth]{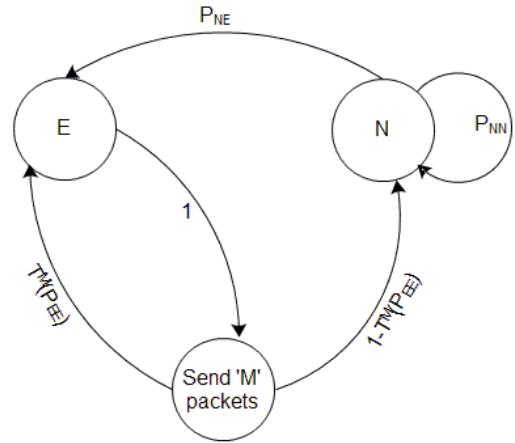}\\
  \caption{Throughput Maximizing Scheme.}\label{fig:3states1}
\end{figure}

We can model the problem by a three state Markov chain as shown in Figure \ref{fig:3states1}:
\\1. Erasure state and the secondary is silent ($E$).
\\2. Non-erasure state and the secondary is silent ($N$).
\\3. Secondary sends $M$ consecutive packets ($S$).
\\When the Markov chain is in the ($N$) state, the primary achieves a throughput of $r_{\rm p}$. When it is in the ($S$) state, the secondary achieves a throughput of $Mr_{\rm s}$ as the system remains in this state for $M$ time slots.

In order to find an expression of the throughput as a function of $M$, we find the stationary distribution of each state of the Markov chain. Let the steady state probability of ($N$), ($E$) and ($S$) states be $P^{ss}_{\rm N}$, $P^{ss}_{\rm E}$ and $P^{ss}_{\rm S}$ respectively, then
\begin{equation}
P^{ss}_{\rm N}=\frac{1-T^M (P_{\rm EE})}{1+2P_{\rm NE}-T^M (P_{\rm EE})}
\end{equation}
\begin{equation}
P^{ss}_{\rm E}=P^{ss}_{\rm S}=\frac{P_{\rm NE}}{1+2P_{\rm NE}-T^M (P_{\rm EE})}
\end{equation}
\noindent where $T^M (P_{\rm EE})$ is the probability of erasure at time slot $t$ given that the state of Markov chain at time slot $t-M$ was erasure. We can find $T^M (P_{\rm EE})$ from the two-state Markov chain:
\begin{equation}
T^M (P_{\rm EE})=\frac{P_{\rm NE}+(P_{\rm EE}-P_{\rm NE})^{(M+1)}\, (1-P_{\rm EE})}{1+P_{\rm NE}-P_{\rm EE}}
\end{equation}
\noindent Recall that we assume a positively correlated channel with $P_{\rm EE}>P_{\rm NE}$.
%It was shown in \cite{Restless Bandit Problems} that for any $P_{\rm EE} \epsilon  [0,1]$,
%$T^k (P_{\rm EE})$ monotonically converges to the steady state distribution $P(E)$
%without any oscillation as $k\rightarrow \infty$.

A closed form expression for the primary throughput $R_{\rm p}$ and the secondary throughput $R_{\rm s}$ can be written as:\\
\begin{equation}
R_{\rm p}=\frac{r_{\rm p}P^{ss}_{\rm N}}{P^{ss}_{\rm N}+P^{ss}_{\rm E}+MP^{ss}_{\rm S}}
\end{equation}
\begin{equation}
R_{\rm s}=\frac{r_{\rm s}MP^{ss}_{\rm S}}{P^{ss}_{\rm N}+P^{ss}_{\rm E}+MP^{ss}_{\rm S}}
\end{equation}
We want to find $M$ that maximizes $Thr(M)=wR_{\rm p}+(1-w)R_{\rm s}$. We can notice that the optimal value of $M$ depends on the weight $w$.
This scheme spans a number of points on the outer bound of the capacity region with the optimal values of $M$ (integer numbers) that maximize the weighted sum throughput. The outer bound of the capacity region here is piecewise linear, and can be achieved by time division multiplexing between the different values of $M$.

Two remarks are in order here:
\\1. Using the properties of the function $V^K(p)$ defined later in Equation (\ref{eqn: itrative bellman}), it can be shown that the optimal strategy is a threshold-based policy on the belief $p_t$, and since the belief $p_t$ is monotonic with $M$, finding the threshold amounts to finding the value of $M$ that maximizes the throughput expression.
\\2. It can be shown that the throughput is a quasi-concave function of $M$, and through some algebraic manipulations, one can arrive at the value of $M$ that maximizes the throughput. This can be shown by subtracting $Thr(M)$ from $Thr(M+1)$. Treating $M$ as a continuous variable, we can show that this difference has only one positive finite root that is greater than or equal to unity. By finding this root, we find the value of the unique optimal $M$.

We will present the details of these proofs in an extended version of this work.

\subsection{Dynamic Programming}

For the three-state channel model presented later and for the case of multiple channels, we may not be able to find a closed-form expression for the throughput. In those cases, we propose to use dynamic programming techniques to arrive at the optimal strategy. In this section we will present the dynamic programming approach to the two-state case.

If we assume an infinite horizon optimization, and through a dynamic programming argument, the state of the system can be fully parameterized by the belief that the channel is in erasure the next time slot, $p$, where we dropped the time dependence. Hence the action taken by the cognitive user depends only on $p$. The belief state $p$ can be updated according to one of the following three cases depending on the action taken by the secondary user and the corresponding outcome. We follow here the notation presented in \cite{Opportunistic Spectrum Access}.

Case 1: Secondary user is silent and a positive acknowledgment (ACK) is received from the primary network. The ACK implies that primary network has been in the non-erasure state, and primary receiver has succeeded in decoding the packet. Therefore, the secondary belief that the channel would be in erasure during the next time slot is
\begin{equation}
L_1(p)=P_{\rm NE}
\end{equation}
where $L_k\left(p\right)$ is the update expression for $p$ for the $k$th case. Each case is a certain combination of secondary decision and observation.

Case 2: Secondary user is silent and a negative acknowledgment (NACK) is received from the primary network. This implies that primary network has been in erasure state and the sent packet has not been delivered successfully. Thus,
\begin{equation}
L_2(p)=P_{\rm EE}
\end{equation}
Case 3: Secondary user transmits. The probability of erasure is updated by the Markovian property as follows
\begin{equation}
L_3(p)=pP_{\rm EE}+\left(1-p\right)P_{\rm NE}
\end{equation}
\noindent We denote the weighted expected instantaneous throughput when the secondary user listens by $G_1\left(p\right)$ which is given by
\begin{equation}
G_1\left(p\right)=wr_{\rm p}\left(1-p\right)
\end{equation}
\noindent For the case of the secondary transmitting, we denote the weighted expected throughput by
\begin{equation}
G_2\left(p\right)=\left(1-w\right)r_{\rm s}
\end{equation}

A greedy scheme would just compare $G_1(p)$ with $G_2(p)$ and if $G_2(p)>G_1(p)$, the secondary user decides to transmit, otherwise it remains silent. The expected instantaneous reward is:
\begin{equation*}
R(p,t) = \left\{ \begin{array}{rcl}
{G_1(p)} & \mbox{if the secondary is silent at time $t$}
 \\ {G_2(p)} & \mbox{if the secondary is active at time $t$}
\end{array}\right.
\end{equation*}

The optimal strategy, on the other hand, takes into account the expected future reward. The optimal strategy is the strategy that maximizes the following discounted reward function \cite{Opportunistic Spectrum Access}
\begin{equation}
E\left\{\sum^{K+t-1}_{n=t} \alpha^{n-t}R\left(p_n,n\right) \hspace{2mm}|\hspace{2mm} p_t=p\right\}
\end{equation}
where $0<\alpha<1$ is a discounting factor and $1 \leq K \leq \infty$ is the control horizon. As $\alpha$ decreases, the secondary user puts more emphasis on its short-term future gains.

Following the definitions in \cite{Opportunistic Spectrum Access}, let $V^K\left(p\right)$ denote the maximum achievable discounted reward function. When $K<\infty$, $V^K\left(p\right)$ satisfies the following Bellman equation \cite{Dynamic Programming}:
\begin{equation}
\label{eqn: itrative bellman}
\begin{split}
&V^K(p)= {\rm max}\left\{wr_{\rm p}(1-p)+\alpha (1-p)V^{K-1}\left(L_1\left(p\right)\right)+\right. \\
& \left. \alpha pV^{K-1}\left(L_2\left(p\right)\right),(1-w)r_{\rm s}+\alpha V^{K-1}(L_3\left(p\right)) \right\}
\end{split}
\end{equation}
where
%\begin{equation}
%V^M(p)= max\left\{ \begin{array}{ll}
%{w*r_p*(1-p)+} & \mbox{}
% \\ {\alpha (1-p)*V^{M-1}(L_1(p))+} & \mbox{}
% \\ {\alpha p*V^{M-1}(L_2(p)),} & \mbox{}
% \\ {(1-w)*r_s+\alpha V^{M-1}(L_3(p))} & \mbox{}
%\end{array}\right.
%\end{equation}
\begin{equation}
V^1(p)= {\rm max} \left\{wr_{\rm p}(1-p), (1-w)r_{\rm s}\right\}
\end{equation}
When $K=\infty$, $V^K(p)=V^{K-1}(p)=V(p)$ which satisfies the following Bellman equation:
\begin{equation}
\label{eqn: steady state bellman}
\begin{split}
&V(p)= {\rm max}\left\{wr_{\rm p}(1-p)+\alpha (1-p)V\left(L_1\left(p\right)\right)+\right. \\
& \left. \alpha pV\left(L_2\left(p\right)\right),(1-w)r_{\rm s}+\alpha V(L_3\left(p\right)) \right\}
\end{split}
\end{equation}
We solve Equation (\ref{eqn: itrative bellman}) iteratively via approximating the value function at a finite number of belief values on a grid (see, for instance, \cite{Dynamic Programming} and \cite{Lovejoy}). The value function is initialized and then (\ref{eqn: itrative bellman}) is used to update it. For $p$ values not belonging to the grid, interpolation or extrapolation is used. After convergence, the secondary terminal decides whether to transmit or listen based on the term that maximizes $V(p)$ at each value of $p$.
%\begin{equation}
%V(p)= max\left\{ \begin{array}{ll}
%{w*r_p*(1-p)+\alpha (1-p)V(L_1(p))+} & \mbox{}
% \\ {\alpha p*V(L_2(p)),} & \mbox{}
% \\ {(1-w)*r_s+\alpha V(L_3(p))} & \mbox{}
%\end{array}\right.
%\end{equation}
%\begin{figure}[!ht]
%  % Requires \usepackage{graphicx}
%  \includegraphics[width=0.8\columnwidth]{DistanceDiff_vd2.eps}\\
%  \caption{simulation}\label{fig:1}
%\end{figure}
\section{Three-state system model}
\label{sec:Three-state system model}
In this section, we extend our previous channel model to a three-state model where the primary channel now follows a three-state Markov chain whose states are named Bad (B), Good (G) and Very good (Vg) with transition matrix $P$ where
\\
\[
P =
\left[ {\begin{array}{ccc}
 P_{\rm BB} & P_{\rm BG} & P_{\rm BVg} \\
 P_{\rm GB} & P_{\rm GG} & P_{\rm GVg} \\
 P_{\rm VgB} & P_{\rm VgG} & P_{\rm VgVg} \\
 \end{array} } \right]
\]
If the secondary is listening, the primary user can deliver its packet if the channel state is G or Vg. But if the secondary is transmitting, the primary transmission success is only in the Vg state. This means that the primary and secondary can both simultaneously transmit successfully in the Vg state.

We can also apply dynamic programming on that system with three channel states to arrive at the optimal decisions for the secondary, whether to transmit or to listen, at any situation to maximize the weighted sum of the primary and secondary throughput.

Here we parameterize the belief state by two parameters $p$ and $q$, where $p$ is the probability that the primary network is in the G state in the next time slot and $q$ is the probability that the primary network is in the Vg state in the next time slot. This implies that the probability that the primary network is in the B state is $(1-p-q)$. After each time slot, depending on the action taken by secondary user and the corresponding feedback, $p$ and $q$ can be updated according to one of the following four cases.
\\Case 1: Secondary user is silent and a NACK is received from the primary network. The NACK during secondary silence implies that primary network has been in B state and, thus, the primary receiver has failed to receive the packet. Therefore, the belief state in the next time slot is:
\begin{equation}
L_1(p)=P_{\rm BG}
\end{equation}
\begin{equation}
L_1(q)=P_{\rm BVg}
\end{equation}
where, as in the two-state case, $L_k(p)$ and $L_k(q)$ are the update expressions for $p$ and $q$, respectively, for the $k$th case.
\\Case 2: Secondary user is silent and an ACK is received from the primary network. Primary network could be in G state with probability $\frac{p}{p+q}$ or Vg state with probability $\frac{q}{p+q}$. The belief state in the next time slot is:
\begin{equation}
L_2(p)=\frac{p}{p+q}P_{\rm GG}+\frac{q}{p+q}P_{\rm VgG}
\end{equation}
\begin{equation}
L_2(q)=\frac{p}{p+q}P_{\rm GVg}+\frac{q}{p+q}P_{\rm VgVg}
\end{equation}
\\Case 3: Secondary user is transmitting and an ACK is received from the primary network. The ACK during secondary activity implies that primary network has been in Vg state. Therefore, the belief state in the next time slot is:
\begin{equation}
L_3(p)=P_{\rm VgG}
\end{equation}
\begin{equation}
L_3(q)=P_{\rm VgVg}
\end{equation}
\\Case 4: Secondary user is transmitting and a NACK is received from the primary network. Primary network could be in G state with probability $\frac{p}{1-q}$ or B state with probability $\frac{1-p-q}{1-q}$. The belief state in the next time slot is:
\begin{equation}
L_4(p)=\frac{p}{1-q}P_{\rm GG}+\frac{1-p-q}{1-q}P_{\rm BG}
\end{equation}
\begin{equation}
L_4(q)=\frac{p}{1-q}P_{\rm GVg}+\frac{1-p-q}{1-q}P_{\rm BVg}
\end{equation}
Let $Q_i(p,q)$, $i=1,2,3,4$, denotes the probability that case $i$ above happens:
\begin{equation}
Q_1(p,q)=1-p-q\\
\end{equation}
\begin{equation}
Q_2(p,q)=p+q\\
\end{equation}
\begin{equation}
Q_3(p,q)=q\\
\end{equation}
\begin{equation}
Q_4(p,q)=1-q\\
\end{equation}

The parameters $p$ and $q$ characterizing the belief state are updated by one of the previous four conditions.
If secondary user is listening, the expected current gain can be calculated as:
\begin{equation}
G_1(p,q)=wr_{\rm p}(p+q)
\end{equation}
But if the secondary user is transmitting, the expected current gain is:
\begin{equation}
G_2(p,q)=(1-w)r_{\rm s}+wr_{\rm p}q
\end{equation}
The expected current reward is:\\
$R(p,q,t) = \left\{ \begin{array}{rcl}
{G_1(p,q)} & \mbox{if the secondary is silent at time $t$}
 \\ {G_2(p,q)} & \mbox{if the secondary is active at time $t$}
\end{array}\right.$
The optimal strategy is the strategy that maximizes the following discounted reward function
\begin{equation}
E\left\{{\sum^{K-1}_{n=0} \alpha^n*R(p_n,q_n,t_n) \hspace{2mm}|\hspace{2mm} p_0=p}\right\}
\end{equation}
$V^K(p,q)$ satisfies the following Bellman equation \cite{Dynamic Programming}:
%\begin{equation}
%V^M(p,q)= max\left\{ \begin{array}{ll}
%{w*r_p*(p+q)+} & \mbox{}
% \\ {\alpha \sum^{}_{i=1,2} Q_i(p,q)V^{M-1}(p_i,q_i),} & \mbox{}
% \\ {(1-w)*r_s+w*r_p*q+} & \mbox{}
% \\ {\alpha \sum^{}_{i=3,4} Q_i(p,q)V^{M-1}(p_i,q_i)} & \mbox{}
%\end{array}\right.
%\end{equation}
\begin{equation}
\begin{split}
&V^K(p,q)= {\rm max}\left\{wr_{\rm p}(p+q)+\alpha \sum^{}_{i=1,2} Q_i(p,q) \right. \\
& \left. V^{K-1}(L_i(p),L_i(q)),~(1-w)r_{\rm s}+wr_{\rm p}q+ \right. \\
& \left. \alpha \sum^{}_{i=3,4} Q_i(p,q)V^{K-1}(L_i(p),L_i(q)) \right\}
\end{split}
\end{equation}
where
%\begin{equation}
%V^1(p,q)= max\left\{ \begin{array}{ll}
%{w*r_p*(p+q),} & \mbox{}
% \\ {(1-w)*r_s+w*r_p*q} & \mbox{}
%\end{array}\right.
%\end{equation}
\begin{equation}
\begin{split}
&V^1(p,q)= {\rm max}\left\{wr_{\rm p}(p+q), (1-w)r_{\rm s}+wr_{\rm p}q\right\}
\end{split}
\end{equation}
When $K=\infty$, $V(p,q)$ denote the maximum achievable discounted reward function.
$V(p,q)$ satisfies the following Bellman equation \cite{Dynamic Programming}:
%\begin{equation}
%V(p,q)= max\left\{ \begin{array}{ll}
%{w*r_p*(p+q)+} & \mbox{}
% \\ {\alpha \sum^{}_{i=1,2} Q_i(p,q)V(p_i,q_i),} & \mbox{}
% \\ {(1-w)*r_s+w*r_p*q+} & \mbox{}
% \\ {\alpha \sum^{}_{i=3,4} Q_i(p,q)V(p_i,q_i)} & \mbox{}
%\end{array}\right.
%\end{equation}
\begin{equation}
\begin{split}
&V(p,q)= {\rm max}\left\{wr_{\rm p}(p+q)+\alpha \sum^{}_{i=1,2} Q_i(p,q)V(L_i(p),L_i(q)),\right. \\
& \left. (1-w)r_{\rm s}+wr_{\rm p}q+\alpha \sum^{}_{i=3,4} Q_i(p,q)V(L_i(p),L_i(q)) \right\}
\end{split}
\end{equation}
\section{Simulation Results}
\label{sec:Simulation Results}
\subsection{Two-State model}
\begin{figure}[!ht]
\centering
  \includegraphics[width=0.9\columnwidth]{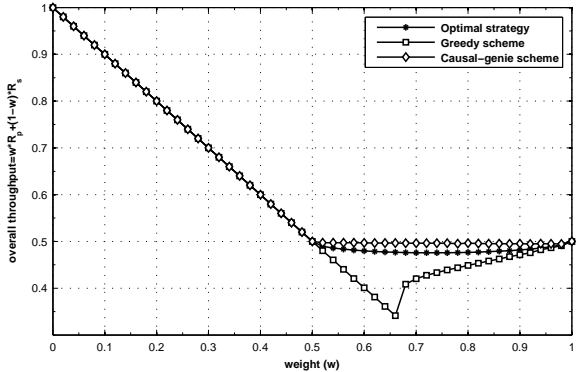}\\
  \caption{Two-state weighted sum throughput.}\label{fig:overall_throughput_2states}
\end{figure}
\begin{figure}[!ht]
\centering
  \includegraphics[width=0.9\columnwidth]{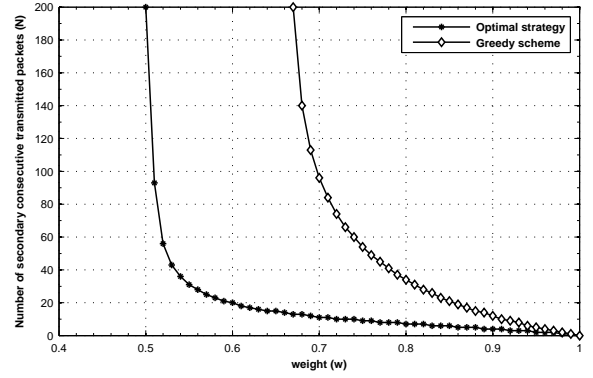}\\
  \caption{Optimal number of consecutive transmitted packets.}\label{fig:M_vs_w}
\end{figure}

For the obtained simulation results, the system parameters are as follows:
$P_{\rm EE}=0.99, P_{\rm NE}=0.01, r_p=1$ and $r_s=1$.
The weighted sum of the primary and secondary throughput is shown in Figure \ref{fig:overall_throughput_2states}. Figure \ref{fig:M_vs_w} shows the optimal values of number of secondary consecutive transmitted packets $M$ versus different values of the weighting factor $w$. We can see from Figure \ref{fig:M_vs_w} that in the greedy scheme, the secondary transmitter transmits always ($M$ is infinite) as long as $w < 0.67$ which explains the sudden change in the overall throughput as $w = 0.67$ in Figure \ref{fig:overall_throughput_2states}. The optimal strategy has this threshold at $w = 0.5$ which means that the optimal strategy benefits more from learning the channel state rather than transmitting to maximize its future reward.

\begin{figure}[!ht]
\centering
  \includegraphics[width=0.9\columnwidth]{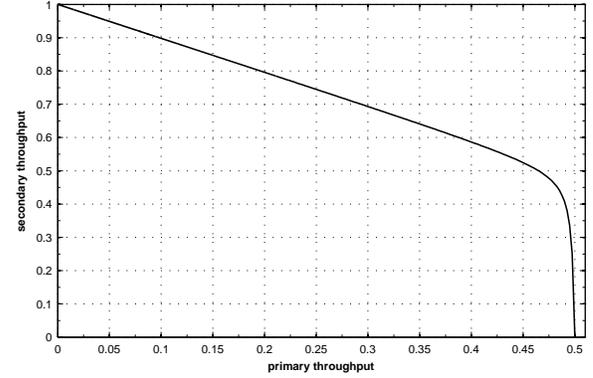}\\
  \caption{primary-secondary rate region.}\label{fig:capacity_region}
\end{figure}

Our proposed scheme spans the boundary of the primary-secondary rate region at number of points where $M$ has an integer value. The piecewise linear connection between these points can be achieved by time division multiplexing between different values of integer $M$. For system parameters $r_p=1, r_s=1$ with the same transition probabilities, the rate region is shown in Figure \ref{fig:capacity_region}.
\subsection{Three-State model}
\begin{figure}[!ht]
\centering
  \includegraphics[width=0.9\columnwidth]{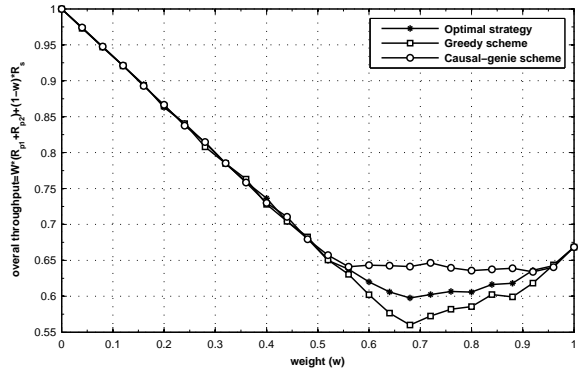}\\
  \caption{Three-state weighted sum throughput.}\label{fig:overall_throughput_3states}
\end{figure}

The system parameters are as follows:$P_{\rm BB}=P_{\rm GG}=P_{\rm VgVg}=0.9, P_{\rm BG}=P_{\rm GB}=P_{\rm VgG}=0.05, r_p=1$ and $r_s=1$. The weighted sum of the primary and secondary throughput is shown in Figure \ref{fig:overall_throughput_3states}.

\section{Conclusion}
\label{sec:Conclusion}
In this paper, the ACK/NACK feedback from the primary receiver is exploited by the secondary transmitter in order to find optimal transmission strategies that maximize the weighted sum of primary and secondary throughput. For the two-state system model, we have derived a closed-form expression of the optimal overall throughput. We have extended the problem to the case of three channel states and used dynamic programming to obtain the optimal secondary policy. Our future work includes the study of multiple primary channels under the assumption of various sensing and access secondary capabilities, as well as the study of multiple secondary users collaborating and competing for transmission opportunities.



%\bibliographystyle{IEEEtran}

\end{document}